\documentclass{aa}  

\usepackage{graphicx}

\usepackage{subfig}

\usepackage{txfonts}

\usepackage{adjustbox}

\usepackage{multirow}

\usepackage{booktabs}

\usepackage{hyperref}

\usepackage{color}

\def\Rsun{{R}_{\sun}}
\def\Lsun{{L}_{\sun}}
\def\Msun{{M}_{\sun}}

\def\diamH{\diameter_\mathrm{H}}
\def\diamK{\diameter_\mathrm{K}}
\def\Bp{B_\mathrm{p}}

\def\chir{\chi^2_\mathrm{r}}

\def\micron{\mu\mathrm{m}}
\def\Teff{T_\mathrm{eff}}

\begin{document} 

   \title{Refined fundamental parameters of Canopus from combined near-IR interferometry and spectral energy distribution \thanks{Based on observations performed at ESO, Chile under program IDs 60.A-9237 and 092.D-0366 for PIONIER, and 084.D-0151 (VISA-CNRS GTO) for AMBER.}}
   
   




\authorrunning{A. Domiciano de Souza et al.}
\titlerunning{Canopus}

   \author{A. Domiciano de Souza \inst{\ref{loca}}\fnmsep\thanks{\email{Armando.Domiciano@oca.eu}}
          \and
          J. Zorec\inst{\ref{sorbiap}} 
          \and
          F. Millour\inst{\ref{loca}}
          \and
          J.-B. Le Bouquin\inst{\ref{ipag}}
          \and 
          A. Spang\inst{\ref{loca}}
          \and
          F. Vakili\inst{\ref{loca}}
          }

   \institute{{Universit{\'e} C{\^o}te d'Azur, Observatoire de la C{\^o}te
    d'Azur, CNRS, Laboratoire Lagrange, France \label{loca}}
    \and
    {Sorbonne Universit{\'e} CNRS, UMR 7095, Institut d'Astrophysique de Paris, F-75014 Paris, France \label{sorbiap}}
    \and
    {Universit{\'e} Grenoble Alpes, CNRS, IPAG, 38000 Grenoble, France \label{ipag}}
             }

   \date{Received ...; accepted ...}

 
  \abstract
   {Canopus, the brightest and closest yellow supergiant to our Solar System, offers a unique laboratory for understanding the physics of evolved massive stars.}
   {We aim at quantitatively exploring a large space of fundamental parameters of Canopus based on the combined analysis of its spectral energy distribution (SED) and optical-IR long baseline interferometry.}
   {We use the most recent high resolution near-IR data from the VLTI focal beam combiners PIONIER (H and K bands) and AMBER (K band), together with precise spectrophotometric measures that cover the SED of Canopus, from the UV to the IR, taken from ground and space observatories.}
   {The accurate and precise PIONIER data allowed us to simultaneously measure the angular diameter and the limb darkening (LD) profile using different analytical laws. We found that the power-law LD, being also in agreement with predictions from stellar atmosphere models, reproduces the interferometric data well. For this model we measured an angular diameter of $7.184 \pm 0.0017 \pm 0.029$ mas and an LD coefficient of $0.1438 \pm 0.0015$, which are respectively $\gtrsim 5$ and $\sim15-25$ more precise than in our previous A\&A paper on Canopus from 2008. 
   From a dedicated analysis of the interferometric data, we also provide new constraints on the putative presence of weak surface inhomogeneities. Additionally, we analyzed the SED in a innovative way by simultaneously fitting the reddening-related parameters and the stellar effective temperature and gravity. We find that a model based on two effective temperatures is much better at reproducing the whole SED, from which we derived several parameters, including a new bolometric flux estimate: $f_{\rm bol} = (59.22\pm2.45)\times10^{-6}$ erg\,cm$^{-2}$\,s$^{-1}$. We were also able to estimate the stellar mass from these measurements, and it is shown to be in agreement with additional predictions from evolutionary models, from which we inferred the age of Canopus as well.}
  {The Canopus angular diameter and LD measured in this work with PIONIER are the most precise to date, with a direct impact on several related fundamental parameters. Moreover, thanks to our joint analysis, we were able to determine a set of fundamental parameters that simultaneously reproduces both high-precision interferometric data and a good quality SED and, at the same time, agrees with stellar evolution models. This refined set of fundamental parameters constitutes a careful balance between the different methodologies used, providing invaluable observationally based constraints to models of stellar structure and evolution, which still present difficulties in simulating stars such as Canopus in detail.}

  
\maketitle

\section{Introduction}


\object{Canopus} (\object{$\alpha$ Car}, \object{HD 45348}) is the brightest star of the southern night sky, and, as such, it has been extensively studied and observed. Its reported spectral type turns around the yellow supergiant class, for example A9~II (SIMBAD) and F0~Ib-II \citep{Jerzykiewicz2000_v50p369}. Canopus is thus a rare, massive \citep[$\sim10\Msun$; e.g.,][]{Jerzykiewicz2000_v50p369, Smiljanic2006_v449p655} yellow supergiant that can be well observed from Earth, providing a unique opportunity to measure its fundamental parameters precisely, which is a key step toward understanding in detail the physical structure and evolution of massive stars beyond the main sequence.

One of the crucial quantities for constraining stellar fundamental parameters is the angular diameter, $\diameter$, of the star. The first direct measure of the angular diameter of Canopus, used by many authors, was made by \citet{HanburyBrown1974_v167p121}, based on the intensity interferometry technique. Several other direct measurements -- notably from optical-IR long baseline interferometry (OLBI) -- and indirect measurements exist in the literature, with reported angular diameters around $\sim6-7$ milliarcseconds (mas). \citet{Cruzalebes2013_v434p437} provide a compilation of many of these angular diameter estimates. It is important to note that a precise estimation of $\diameter$ from high angular resolution techniques is intimately related to the estimation of the center-to-limb intensity profile -- limb darkening (LD) -- at the spectral domain of the observation. These LD estimations also provide important constraints on the physical structure of the stellar atmosphere, which depend on several physical parameters (e.g., mass, temperature, gravity, chemical composition). These observational constraints are particularly important for yellow supergiants such as Canopus because models of stellar structure and evolution still have difficulties in simulating stars in this region of the Hertzsprung-Russell (HR) diagram. A precise $\diameter$ measure, including the LD effect, leads to a more realistic estimate of the stellar radius from $R=0.5\diameter d$, where $d$ is the distance to the star. For Canopus one has $d=94.8\pm5.0$~pc, corresponding to the Hipparcos parallax $\pi=10.55\pm0.56$~mas \citep{vanLeeuwen2007_v474p653}. No Gaia distance is available for a star as bright as Canopus.

The OLBI studies of Canopus provide direct constraints not only on its fundamental parameters but also on the presence of surface features since this star is spatially well resolved by modern interferometers. Indeed, Canopus is known to exhibit several degrees of activity (e.g., high energy UV and X-ray emission) and temporal variations, the origins of which are only partly understood and can induce photospheric structures (e.g., spots). These issues related to the activity on Canopus have been largely addressed in the past, for example by \citet{Rakos1977_v56p453}, \citet{Vaiana1981_v245p163}, \citet{Weiss1986_v160p243}, \citet{Dupree2005_v622p629}, \citet{Bychkov2009_v394p1338}, and \citet{Ayres2018_v854p95}, among several others. 


In a previous interferometric study of Canopus \citep{DomicianodeSouza2008_v489pL5}, we used observations from the near-IR beam combiner AMBER \citep{Petrov2007_v464p1} of the ESO Very Large Telescope Interferometer \citep[VLTI;][]{Haguenauer2010_v7734p1} to estimate its angular diameter and LD. This first work also showed the presence of photospheric structures, but no strong constraints could be imposed on these structures because of the limited quantity and quality (S/N and absolute visibility calibration) of the available data.

These findings motivated us to revisit Canopus (i) by using more precise interferometric data from the near-IR beam combiner PIONIER/VLTI \citep{LeBouquin2011_v535pA67} and (ii) by rebuilding a new spectral energy distribution (SED) with recent and precise spectrophotometry in the visible, combined with previously reported observations on the other spectral domains. We also use new AMBER data as a cross-check of the results from PIONIER, although they lead to lower-precision measurements. 

The observations and analysis of the near-IR interferometric and SED data are described in Sects.~\ref{vlti_analysis} and \ref{sed_analysis}, respectively. In Sect.~\ref{discussion} we discuss our results, comparing them with several other previous results and with stellar atmosphere models. We also investigate and discuss the presence of surface structures based on both model fitting and image reconstruction. The conclusions of our work are presented in Sect.~\ref{conclusions}.


\section{Interferometric analysis \label{vlti_analysis}}

\subsection{Near-IR VLTI observations and data reduction \label{vlti_obs_datared}}

\subsubsection{PIONIER data \label{pionier}}



Because of their good quality, the main interferometric observations of Canopus considered in this work are those obtained over four nights (one in 2010 and three in 2014) with the beam-combiner instrument PIONIER. PIONIER uses integrated optics technology and can simultaneously combine the light beams of the four 1.8 m Auxiliary Telescopes (ATs) of the VLTI. 

The raw data of Canopus and of the calibrator stars (mainly HD49517, HD54792, and HD39640) were reduced using the standard PIONIER pipeline \textit{pndrs} \citep{LeBouquin2011_v535pA67}. For each observation of Canopus, PIONIER provides six calibrated squared visibilities, $V^2$, and three independent closure phases, $CP$. 

\begin{figure*}[ht!]
\centering
\subfloat[]{\includegraphics[clip, trim=2cm 0cm 2cm 0cm, width=0.48\textwidth]{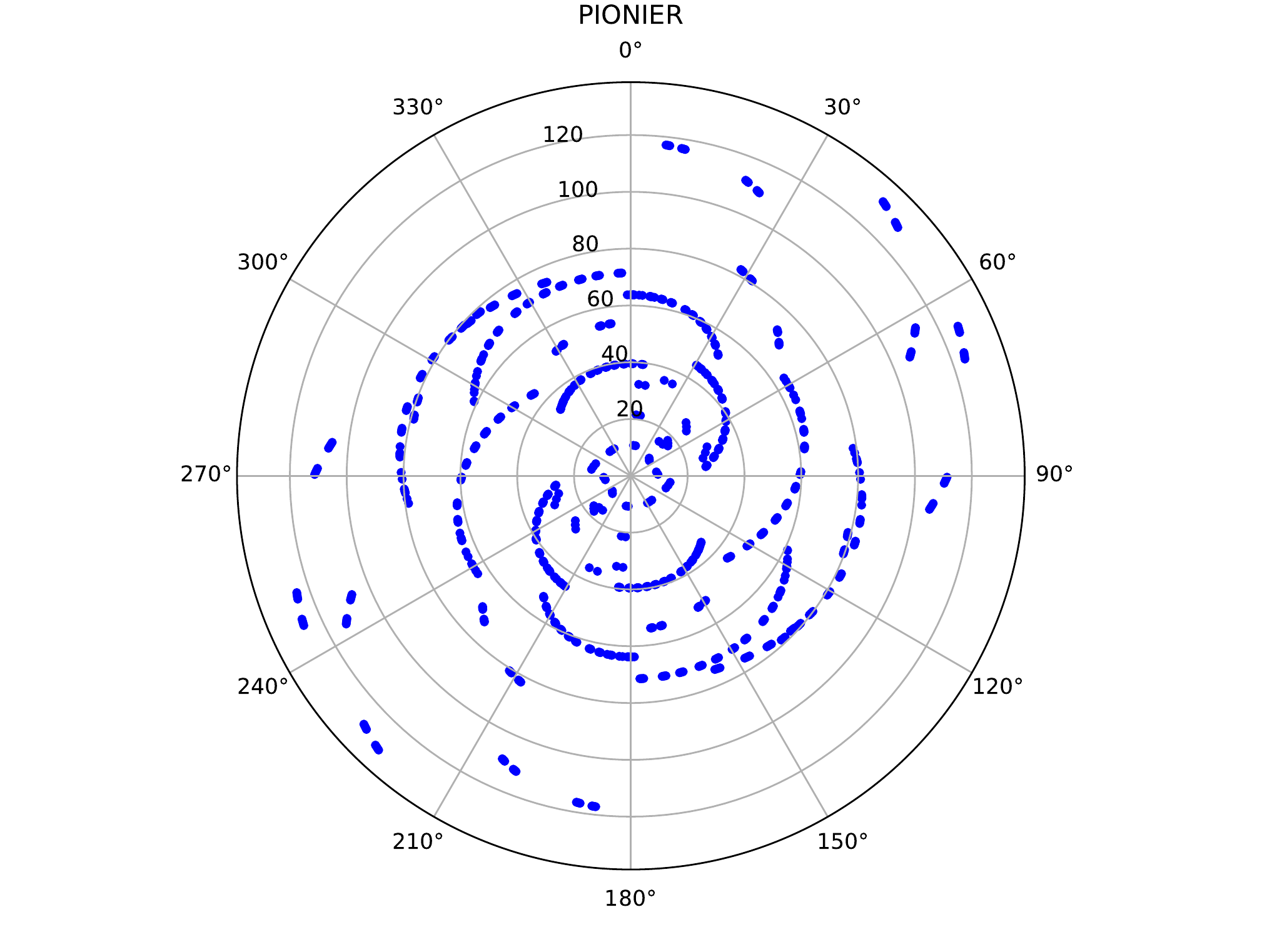}}%
\qquad
\subfloat[]{\includegraphics[clip, trim=2cm 0cm 2cm 0cm, width=0.48\textwidth]{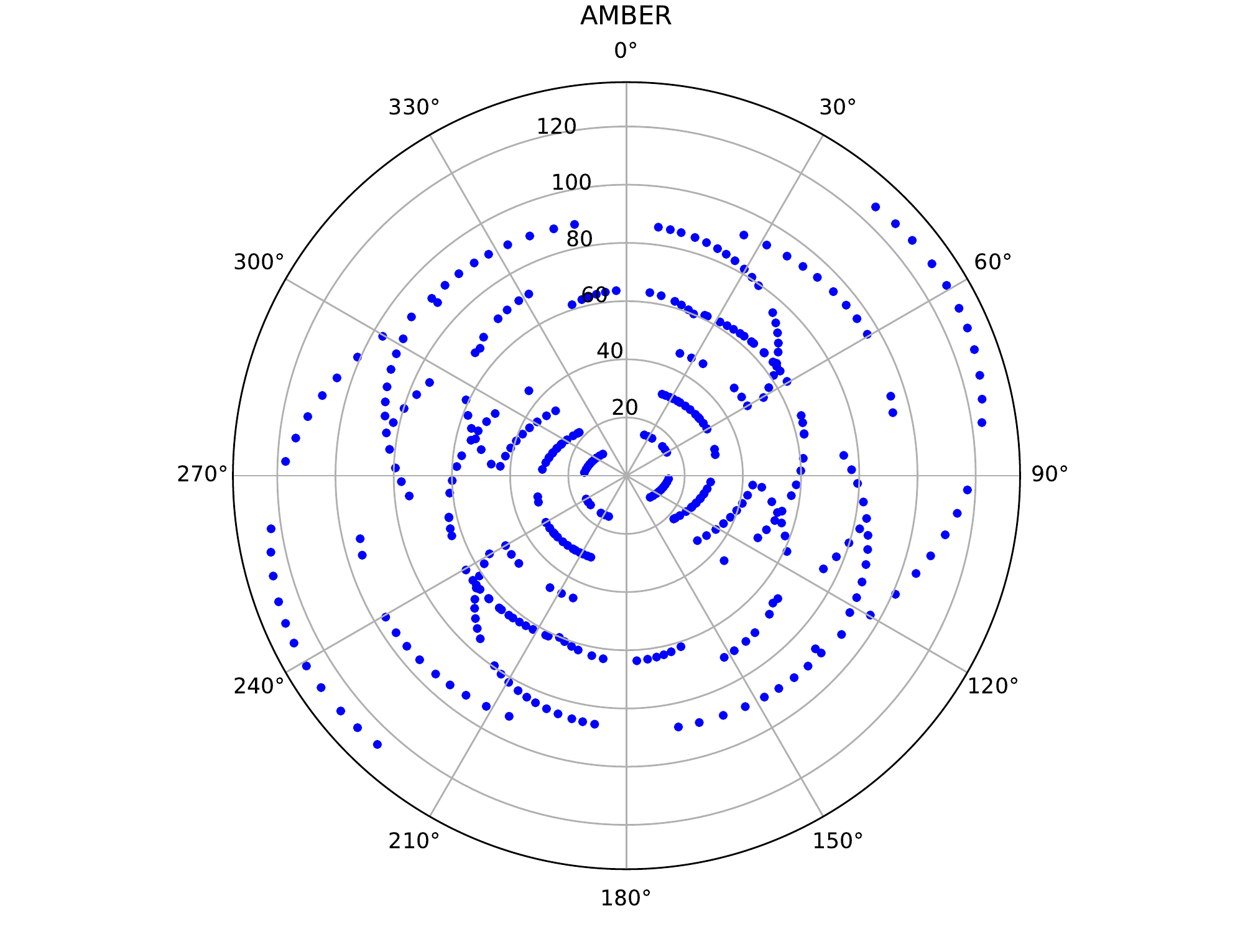}}
\caption{\label{fig:uv_planes} Fourier or $uv$-plane coverage for the VLTI observations of Canopus obtained with PIONIER (a) and AMBER (b) using different combinations of ATs. The plots are given in polar coordinates so that the values of the projected baseline lengths, $\Bp$ (circles, with values in meters), and position angles, PA (radial lines, given in degree), are explicitly indicated.}
\end{figure*}

These interferometric observables cover the H band (around $1.65\,\micron$) with a spectral resolution of $\lambda/\delta\lambda \approx 40$: six spectral channels between $\sim1.6\,\micron$ and $\sim1.8\,\micron$ in 2010 and three spectral channels between $\sim1.6\,\micron$ and $\sim1.7\,\micron$ in 2014. Reduced data, along with additional information on the observations, are available at the \texttt{OiDB} web service offered by the Jean-Marie Mariotti Center (JMMC), France. A summary of these PIONIER observations is given in Table~\ref{tab:log_obs_vlti}, and the corresponding $uv$ plane (or Fourier plane) is shown in Fig.~\ref{fig:uv_planes}a.


%
\begin{table}[]
\centering
\caption{\label{tab:log_obs_vlti} Log of the PIONIER and AMBER observations of Canopus considered in this work. }
\begin{tabular}{ccc}
\toprule
\textbf{Date} & \multicolumn{1}{c}{\textbf{Number of}}   & \textbf{AT}  \\
\textbf{(YYYY-MM-DD)}  & \multicolumn{1}{c}{\textbf{data files}} & \textbf{configuration}  \\
\midrule
\multicolumn{3}{c}{\textbf{PIONIER/VLTI (H band)}} \\
\midrule
2010-10-31  &  5 & D0-G1-H0-I1  \\
2014-01-12  &  5 & A1-B2-C1-D0  \\
2014-02-02  &  4 & A1-G1-J3-K0  \\
2014-02-05  & 12 & D0-G1-H0-I1  \\
\midrule
\multicolumn{3}{c}{\textbf{AMBER/VLTI (H and K bands)}} \\
\midrule
2009-12-21  & 12 & D0-G1-H0  \\
2009-12-22  & 16 & D0-H0-K0  \\
2009-12-24  & 20 & A0-G1-K0  \\
2009-12-28  & 15 & H0-G0-E0  \\
\bottomrule
\end{tabular}
\tablefoot{
Observations were recorded with several AT configurations on the VLTI, indicated in Col. 3. The corresponding $uv$-plane coverage is shown in Fig.~\ref{fig:uv_planes}.
}
\end{table}

\subsubsection{AMBER data \label{amber}}

We observed Canopus in 2009 with the near-IR beam-combiner AMBER/VLTI, using three ATs simultaneously. The observations were performed in LR-JHK mode (spectral resolution $\lambda/\delta\lambda\approx30$), within the framework of a VLT Interferometric Sub-Array (VISA)-CNRS (France) guaranteed observation time (GTO) program. The raw data were reduced with \textit{amdlib} \citep[version~3.0.9;][]{Chelli2009_v502p705, Tatulli2007_v464p29}. To improve the signal-to-noise ratio (S/N) of the reduced data, we adopted a fringe frame selection rule with a fringe S/N threshold of $>1$ and an absolute optical path difference (OPD) threshold of $<20\,\micron$. 


The observations of Canopus were interspersed with observations of calibration stars ($\delta$ Phe, HD40808), allowing calibrated observables to be computed: three $V^2$ and  one $CP$ per observation, simultaneously in the H and K bands. The J band was not considered here because the data are of too low quality and this spectral band is not validated by ESO. A summary of these AMBER observations is given in Table~\ref{tab:log_obs_vlti}, and the corresponding $uv$ plane is shown in Fig.~\ref{fig:uv_planes}b. 

We decided not to include in this work the 2007 AMBER data analyzed in our previous work \citep{DomicianodeSouza2008_v489pL5} because they were recorded with a different, lower quality detector and have a distinct spectral calibration. Compared to the 2009 data, the reduction of the 2007 data is thus trickier and less precise, and also results in observables with a distinct spatial frequency systematic shift. Moreover, the 2009 observations are of a higher quality and are more numerous, with a much larger $uv$-plane coverage.


\subsubsection{Spatial frequency precision \label{sfreq_precision}}

To achieve precise interferometric measurements of angular sizes, it is important to have a good calibration of the spatial frequency, $\Bp / \lambda$, where $\Bp(=|\vec{\Bp}|)$ is the interferometric baseline length projected onto the target and $\lambda$ is the effective wavelength of the observation. In a given observation, these quantities are generally affected by systematical deviations that directly impact the measured angular sizes. 

Following the analysis of \citet{Kervella2017_v597pA137}, we consider in this work that the PIONIER measurements of angular diameters are affected by a systematic uncertainty of 0.41\% associated with the instrumental wavelength calibration scale. For the angular diameters measured with AMBER, following \citet{DomicianodeSouza2008_v489pL5}, we consider a systematic uncertainty of one wavelength-bin pixel, which translates to systematic angular diameter uncertainties of $\sim2.0\%$ and $\sim1.5\%$ in the H and K bands, respectively. Uncertainties in the baseline length can be neglected, as discussed by \citet{Kervella2017_v597pA137}. 



\subsection{Diameter and limb darkening from interferometry \label{interf_emcee_results}}

We sought to use the interferometric data of Canopus to measure its angular diameter through a model-fitting procedure. Since interferometry is sensitive to the intensity distribution of the target, it is important to consider the LD effect for a more precise estimation of both the angular diameter ($\diameter$) and the LD itself.

The near-IR VLTI observations of Canopus considered in this work provide visibility data up to the third visibility lobe, as we show in the following. As discussed by \citet{Kervella2017_v597pA137}, this allows us to unambiguously constrain analytical intensity-profile models with up to one LD parameter.

\subsubsection{Limb-darkening models \label{ld_models}}

We investigated Canopus with three radial (1D) intensity profile models, which correspond to different LD laws with zero or one parameter: (i) the classical uniform disk (UD), (ii) the linear LD, which is commonly used, and (iii) the power-law LD, which has been shown to be a realistic model in many cases \citep[e.g.,][]{Hestroffer1997_v327p199, Kervella2017_v597pA137}. 

The complex visibility, and corresponding interferometric observables, associated with these three LD laws can be analytically computed with a Hankel transform of the intensity profiles. These analytical functions are given in Table~\ref{tab:ld_models} for the reader's convenience, although they are well known and can be found in the literature \citep[e.g.,][]{HanburyBrown1974_v167p475, Hestroffer1997_v327p199}. We note that all three of these functions are special cases of a more general form for the LD law, as shown in Appendix~\ref{app:analytical_vis}.

Moreover, we included the bandwidth smearing effect in our LD model calculations of the visibility amplitudes, as explained in Appendix~\ref{band_smear}. This is necessary because Canopus is well resolved by the VLTI (up to the third visibility lobe) and because the PIONIER and AMBER observations were performed in low resolution mode, so different fringe contrasts are mixed together in each spectral bin. This effect is particularly important in the vicinity of the visibility minima.

\begin{table*}[]
\centering
\caption{\label{tab:ld_models} Limb darkening (LD) models used to interpret the interferometric observations of Canopus.}
\begin{tabular}{cccc}
\toprule

\textbf{LD model} & \textbf{Intensity profile} & \multirow{2}{*}{$\xrightarrow[\text{transform}]{\text{Hankel}}$}  & \textbf{Complex visibility function} \\
\textbf{name} & $I{_\lambda}(\mu)/I{_\lambda}(1)$ &   &$V(z)$, where $z=\pi \diameter \Bp / \lambda$ \\

\midrule
Uniform disk (UD)  & 1 & $\rightarrow$ &   $\displaystyle 2 \frac{J_1(z)}{z}$ \\
\midrule
Linear             & $\displaystyle 1-{u_\lambda}(1-\mu)$ & $\rightarrow$ &   $\displaystyle \frac{(1-{u_\lambda})J_1(z)/z + {u_\lambda} \sqrt{\pi/2} J_{1.5}(z)/z^{1.5}}{(1-{u_\lambda})/2+{u_\lambda}/3}$ \\
\midrule
Power law          & $\displaystyle \mu^{\alpha_\lambda}$ & $\rightarrow$ &   $\displaystyle {\nu_\lambda}  \Gamma({\nu_\lambda}) 2^{\nu_\lambda} \frac{J_{\nu_\lambda}(z)}{z^{\nu_\lambda}}$, where ${\nu_\lambda}=\frac{{\alpha_\lambda}}{2}+1$ \\
\bottomrule 

\end{tabular}
\tablefoot{
Equations for the intensity profiles, $I(\mu)$, and the corresponding complex visibilities, $V$, are given. In these equations, $\mu (= \cos(\theta))$ is the cosine of the angle $\theta$ between the direction perpendicular to the stellar surface and the direction toward a distant observer. For a line-of-sight starting at the center and moving toward the limb of the stellar disk, $\mu$ varies between 1 and 0. Also, $\Gamma$ is the gamma function and $J_{\nu_\lambda}(z)$ is the Bessel function of first kind and of order ${\nu_\lambda}$. The dimensionless variable $z$ is given by $\pi \diameter \Bp / \lambda$, where $\Bp(=|\vec{\Bp}|)$ is the interferometric baseline length projected onto the target, $\lambda$ is the effective wavelength of the observation, and $\diameter$ is the stellar angular diameter for the considered LD model. In an interferometric study of a given star, as is the case here, one generally seeks to observe at several spatial frequencies for a good $uv$ coverage, with observations performed on many different realizations of the vector $\vec{\Bp} / \lambda =(u,v),$ as in Fig.~\ref{fig:uv_planes}.
}
\end{table*}

\subsubsection{MCMC model fitting to PIONIER data \label{emcee_res_pionier}}

The three LD models described in the previous section were fitted to PIONIER's $V^2$ and $CP$ data with the \textit{emcee}\footnote{\url{https://emcee.readthedocs.io/en/v3.0.2/}} Python package \citep{Foreman-Mackey2013_v125p306}. This package, largely used in astronomical data analysis, is an implementation of the Goodman \& Weare's affine invariant Markov chain Monte Carlo (MCMC) ensemble sampler \citep{Goodman2010_v5p65}. Because the LD models are analytical (see Table~\ref{tab:ld_models}) and thus fast to compute, the model fitting could be performed with many walkers (1000). For the burn-in and final phases, we adopted, respectively, $200-400$ (depending on the model) and $100$ iterations. Each model fitting takes $\sim1-4$ hours on a standard laptop from 2020, using two to four simultaneous central processing units.

\begin{figure}[ht!]
\centerline{\includegraphics[clip, trim=5.0cm 2.2cm 4.2cm 1.5cm, width=\hsize]{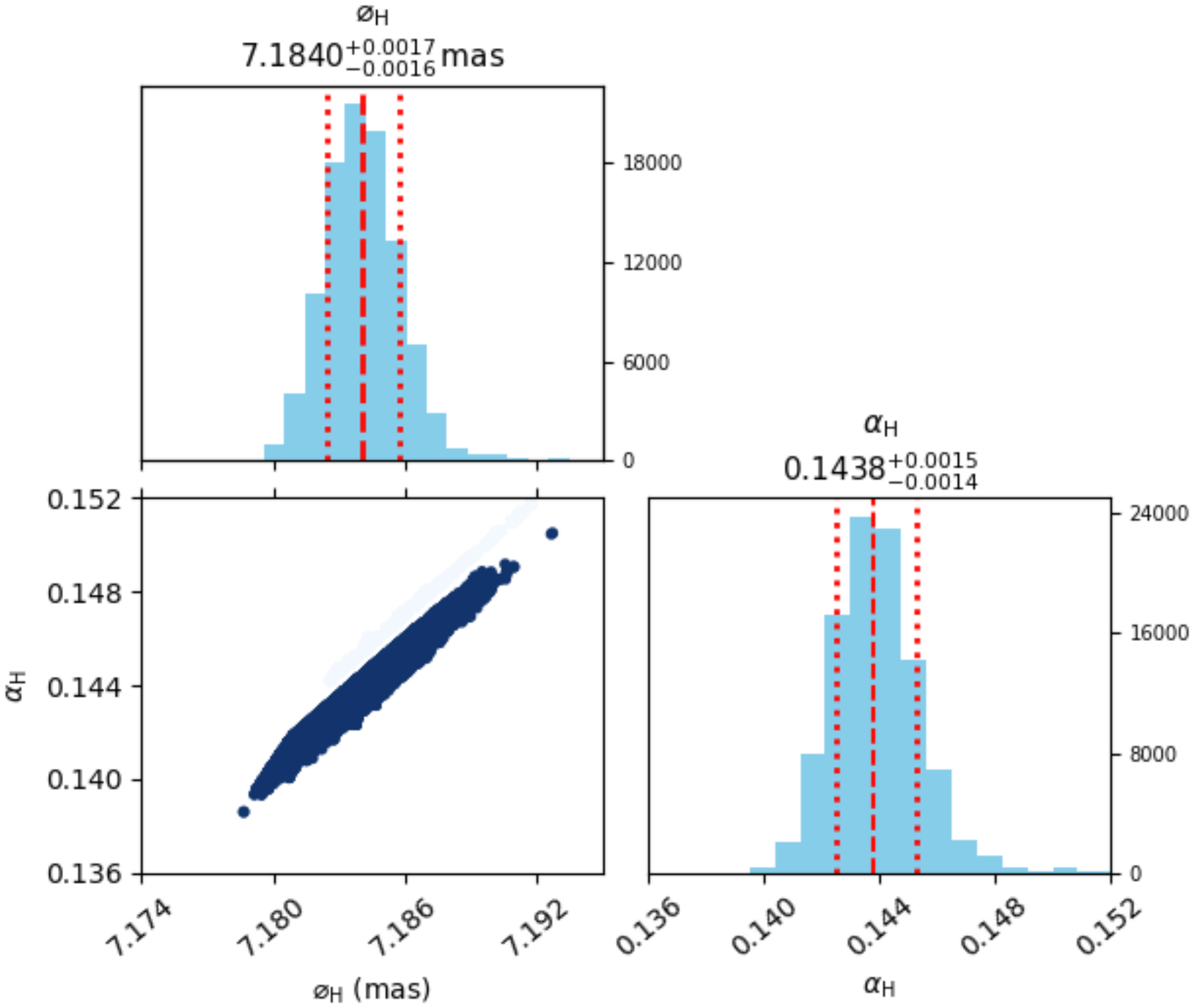}}
\caption{\label{fig:corner_best_fit_power_law_PIONIER} Histograms and correlation plot of model parameters (angular diameter H and $\alpha_\mathrm{H}$) determined from the \textit{emcee} fit of a power-law LD, including bandwidth smearing, to Canopus PIONIER data ($V^2$ and $CP$). The corresponding medians and 16\% and 84\% percentiles are indicated (see Table~\ref{tab:emcee_results_pionier}). In the correlation plot, lower chi-square values correspond to darker symbols.}
\end{figure}

The parameters ($\diameter$ and LD coefficient) measured from the \textit{emcee} fit of the three LD models adopted in this work are summarized in Table~\ref{tab:emcee_results_pionier}, together with the corresponding $\chi^2$. The PIONIER data are poorly represented by the UD model but are reproduced much better by both the linear and the power-law LD models, in particular when considering bandwidth smearing. Anticipating the discussion in Sect.~\ref{compare_Neilson_SAtlas}, where these fitted analytical models are compared to LD profiles from stellar atmosphere models, we choose hereafter the power-law LD as our reference best model to represent Canopus (values in boldface in Table~\ref{tab:emcee_results_pionier}). The values shown in this table are directly computed from histograms built from samples of the posterior probability density function (PDF) provided by \textit{emcee}. As an example, we give in Fig.~\ref{fig:corner_best_fit_power_law_PIONIER} the so-called corner plot built from a PDF sample corresponding to our reference best-fit power-law model. 

\begin{figure*}[ht]
\centerline{\includegraphics[clip, trim=0.8cm 0.7cm 5.2cm 0.5cm, width=\hsize]{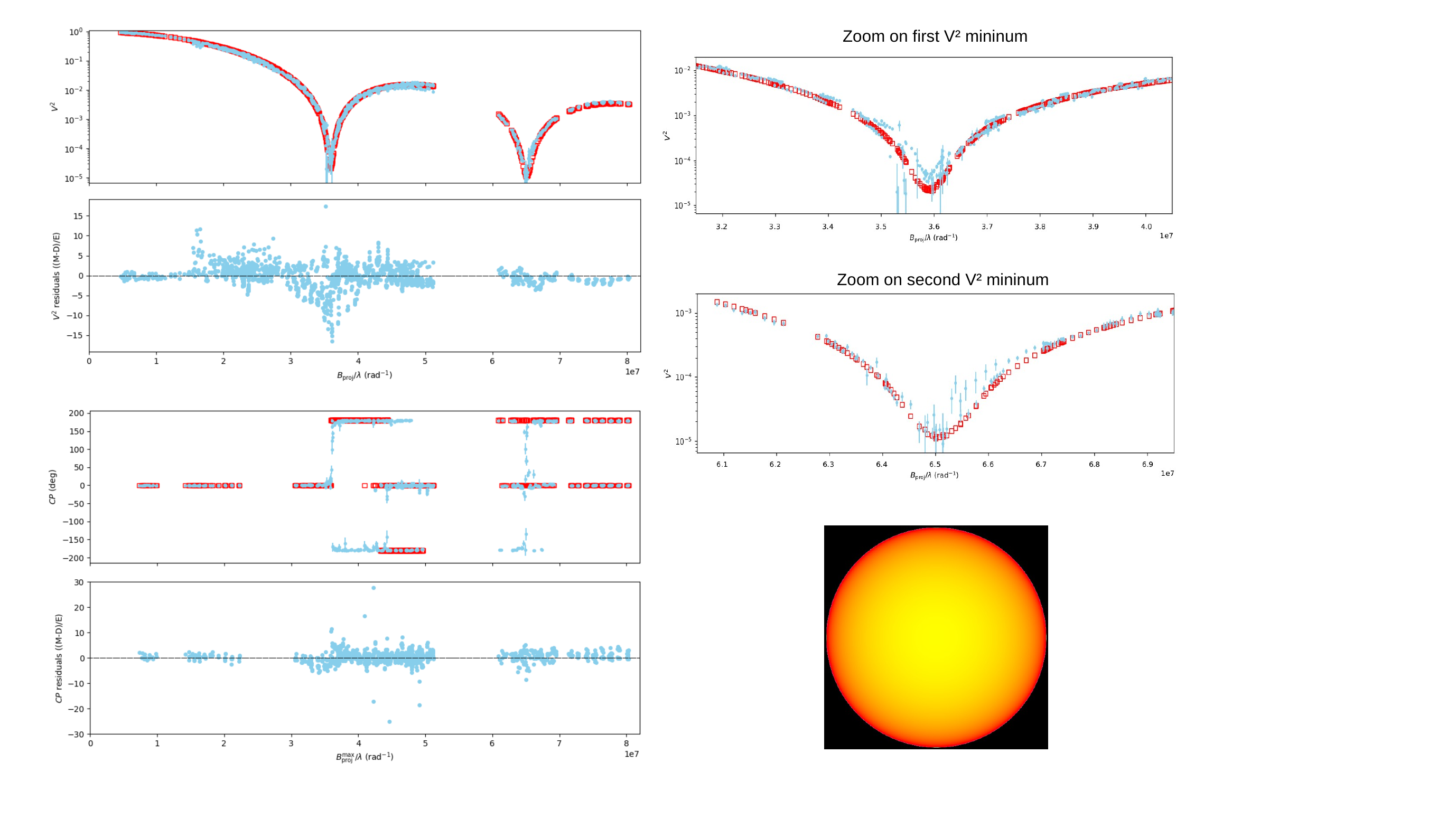}}
\caption{\label{fig:best_fit_power_law_PIONIER} Canopus calibrated squared visibilities, $V^2$ ($\log$ scale), closure phases, $CP$, and errors as a function of the spatial frequency, observed with VLTI/PIONIER (light blue). The red squares correspond to the best-fit-model observables obtained from an MCMC fit on the PIONIER data using a power-law LD and including bandwidth smearing. The residuals of the fit are also shown; the fit details and measured parameters are given in Sect.~\ref{vlti_analysis} and in Table~\ref{tab:emcee_results_pionier}. The two upper-right figures show a zoom-in on the first two $V^2$ minima, both of which are well sampled by PIONIER. The bottom-right image illustrates Canopus as seen in the H band according to our best-fit power-law LD model.}
\end{figure*}

Figure~\ref{fig:best_fit_power_law_PIONIER} shows a comparison of this best-fit model to the $V^2$ and $CP$ measured on Canopus with PIONIER. The corresponding H-band intensity map is also shown in the figure. The observations are well reproduced by the power-law LD model, at all spatial frequencies, up to the middle of the third visibility lobe. In particular, the zoomed-in plots show that the $V^2$ observations at the two first minima are relatively well reproduced by the model, thanks to the inclusion of bandwidth smearing. The larger residual dispersion seen in these $V^2$ minima regions are caused by the poor $V^2$ S/N resulting from the low fringe contrast. This explains why the reduced chi-squares, $\chir$, are sensibly higher than 1 even though the global model fit and residuals are well behaved (for example, without any suspicious residual trend). Indeed, somewhat high $\chir$ values are not uncommon in interferometric works with precise observations that include $V^2$ data \citep[e.g.,][]{Kervella2017_v597pA137}. In any case, to be on the safe side, we verified that our results remain nearly unchanged (compatible within errors) and that the $\chir$ becomes significantly and progressively smaller as we gradually and artificially increase the uncertainties of the $V^2$ data close to the two minima before doing the model fit. In the limit (and unrealistic) case where the $V^2$ uncertainties of the data close to the $V^2$ minima are set to 1, the corresponding $\chir$ on $V^2$ is divided by $\simeq2$, while the parameter values are nearly unchanged and the parameter statistical errors only increase by a factor of $\simeq1.5$.

The results of the MCMC fit of the power-law LD model to the AMBER H and K data are given in Appendix~\ref{best_fit_power_law_AMBER}. In this appendix it is shown that the measured parameters are compatible with those derived from the PIONIER data analysis but less precise because the AMBER data present a much higher dispersion and a lower S/N. In the following, we thus consider the power-law LD parameters obtained from the PIONIER data analysis as the main results of our interferometric study of Canopus.



%
\begin{table*}[]
\centering
\caption{\label{tab:emcee_results_pionier} Best-fit results for three LD models fitted to the Canopus PIONIER data.}
\begin{tabular}{*{5}{c}}
\toprule
\multicolumn{5}{c}{\textbf{PIONIER (H band)}} \\
\toprule
& \multicolumn{2}{c}{\textbf{$V$ model from Table~\ref{tab:ld_models}}} & \multicolumn{2}{c}{\textbf{$V$ model from Table~\ref{tab:ld_models} with bandwidth smearing}} \\
\cmidrule[1pt](lr){2-3}
\cmidrule[1pt](lr){4-5}
\textbf{Model} & \textbf{Parameters} & $\chir$ (total/$V^2$/$CP$)  & \textbf{Parameters} & $\chir$ (total/$V^2$/$CP$) \\
\midrule
UD & $\diamH=7.020 \pm 0.0003 \pm 0.029$ mas  & 18.0/12.0/6.0 & $\diamH=7.019 \pm 0.0001 \pm 0.029$ mas & 17.1/11.1/6.0 \\
\midrule
Linear & $\diamH=7.147 \pm 0.0011 \pm 0.029$ & 8.8/6.2/2.6 & $\diamH=7.146 \pm 0.0014 \pm 0.029$ mas & 7.9/5.3/2.6 \\
  LD   & $u_\mathrm{H}= 0.1793 \pm 0.0013$ & & $u_\mathrm{H}=0.1795 \pm 0.0017$ & \\
\specialrule{.2em}{.1em}{.1em} 
\textbf{Power} & $\diamH=7.186 \pm 0.0017 \pm 0.029$ & 8.9/6.3/2.6 & $\mathbf{\diamH=7.184 \pm 0.0017 \pm 0.029}$ \textbf{mas} & 7.9/5.3/2.6 \\
  \textbf{law LD}  & $\alpha_\mathrm{H}=0.1441 \pm 0.0014$ &  & $\mathbf{\alpha_\mathrm{H}=0.1438 \pm 0.0015}$ &  \\
\specialrule{.2em}{.1em}{.1em} 
\end{tabular}
\tablefoot{
Best-fit results for three LD models (defined in Table~\ref{tab:ld_models}) fitted to the Canopus PIONIER data (2698 data points: 1639 squared visibilities, $V^2$, and 1059 closure phases, $CP$). The best-fit parameter values correspond to medians obtained from the histograms provided by the MCMC fitting tool \textit{emcee}. The statistical uncertainties (second values), also computed from the histograms, correspond to the commonly adopted 16\% and 84\% percentiles ($\pm \sigma$ in the 68\% rule). The third values given for the angular diameters $\diamH$ are the systematical errors associated with the instrumental wavelength calibration of PIONIER (see Sect.~\ref{sfreq_precision}). We give the results obtained with (right) and without (left) considering the bandwidth smearing effect. We also provide the reduced chi-squares, $\chir$, for the whole data set and for the $V^2$ and $CP$ data alone, allowing the fit quality of the different models to be compared. We accounted for phase wrapping in $\chi^2$ computations for $CP$. The parameters for the power-law LD with bandwidth smearing are indicated in boldface since this is the reference model in this work, better representing Canopus data (see text). By adopting the Hipparcos distance, this reference angular diameter results in a radius of $R=0.5 \diamH d=73.2 \pm 3.9 \Rsun$ for Canopus.
}
\end{table*}

\section{SED analysis \label{sed_analysis}}



According to the Gaia-Two Micron All Sky Survey (2MASS) 3D maps of Galactic interstellar dust within 3~kpc of Earth \citep{lall2019}, Canopus is situated in a rather dustless region. Indeed, in the STILISM\footnote{STructuring by Inversion the Local Interstellar Medium; \newline \url{https://stilism.obspm.fr/}} maps \citep{lall2014,capit2017} we read a color excess (reddening) of $E(B-V)=0.001\pm0.020$~mag at the distance of Canopus. This corresponds to an interstellar total visual extinction of $A_{\rm V}=0.0031$~mag, for a visual extinction to reddening ratio of $R_{\rm V}=A_{\rm V}/E(B-V)=3.1$. 

Higher reddening values are estimated by \citet{Kovtyukh2008_v389p1336} ($E(B-V)=0.016 \pm0.050$~mag) and \citet{Gontcharov2018_v2354p} ($E(B-V)=0.07$~mag, with $R_{\rm V}=3.31$). Finally, using the \textit{dustmaps} Python package \citep{Green2019_v887p93} and the NASA/IPAC infrared science archive (IRSA) service for Galactic dust reddening, one finds even higher values: $E(B-V)=0.0873\pm0.0006$~mag \citep{Schlegel1998_v500p525} and $E(B-V)=0.0750\pm0.0005$~mag \citep{Schlafly2011_v737p103}. The iteration procedure used in Sect.~\ref{zorec_params_from_sed} to estimate several fundamental parameters from model fitting to the SED indeed required $E(B-V)=0.0874$~mag and $R_{\rm V}=3.1$, a value that can imply a possible presence of "distant circumstellar shells from earlier epochs of mass loss," as noted by \citet{Ayres2011_v738p120}.

\subsection{Setting the SED \label{sed_obs}}

Because no clear indications of strong photometric variability have been reported for Canopus (see the discussion in Sect.~\ref{activity_spots}), we built its SED by combining multispectral absolute flux and magnitude data obtained at distinct epochs.

The UV spectral range from 1429 \AA\ to 3300 \AA\ has been established using the compilation of absolute spectrophotometric data in the Ultraviolet Bright-Star Catalog \citep{jamar1976} obtained with the Thor-Delta 1A (TD1) satellite, the Astronomical Netherlands Satellite (ANS) UV photometry of point sources \citep{wesse1982}, and the low resolution  International Ultraviolet Explorer (IUE) spectra calibrated in absolute fluxes from the IUE Newly Extracted Spectra (INES) archive data server through the Strasbourg astronomical Data Center (CDS). The calibration of the TD1 satellite ensures flux uncertainties within 11~\% for the short wavelength channel, from 11 to 17~\% for the medium wavelength channel, and 19~\% for the long wavelength channel. The ANS data are given with uncertainties of 20~\% for the absolute internal calibration and less than 10~\% for the relative calibration. The TD1 and ANS fluxes were mixed and considered with their own uncertainties. The IUE data are highly noisy and were used here only for comparison. The visible spectral range of Canopus from 3300 \AA\ to 1 $\mu$m is represented with the spectrophotometric data from \citet{Krisciunas2017_v129p054504}, ensured within 5~\% uncertainty.

The near- and mid-IR spectral domain was built with fluxes from (i) the 2MASS All-Sky Catalog of Point Sources \citep{cutri2003} given with uncertainties that range from 5 to 10~\%, (ii) the Infrared Astronomical Satellite (IRAS) Point Source Catalog \citep{Helou1988}, (iii) data from the Infrared Space Observatory (ISO) Short Wavelength Spectrometer fluxes \citep{krae2002}, where fluxes have uncertainties on the order of 4 to 5\%, and (iv) IR fluxes from the AKARI/Far-Infrared Surveyor (FIS) All-Sky Survey Point Source Catalogs \citep{mura2007,yama2010} that give data within 1 to 4\% uncertainty. Since ISO fluxes are step-like and discontinuous from 2.36 to 10~$\mu$m, we searched for a matching compromise with the J, H, and K fluxes from 2MASS.

\begin{figure}[t!]
\centerline{\includegraphics[width=\hsize, scale=1.0]{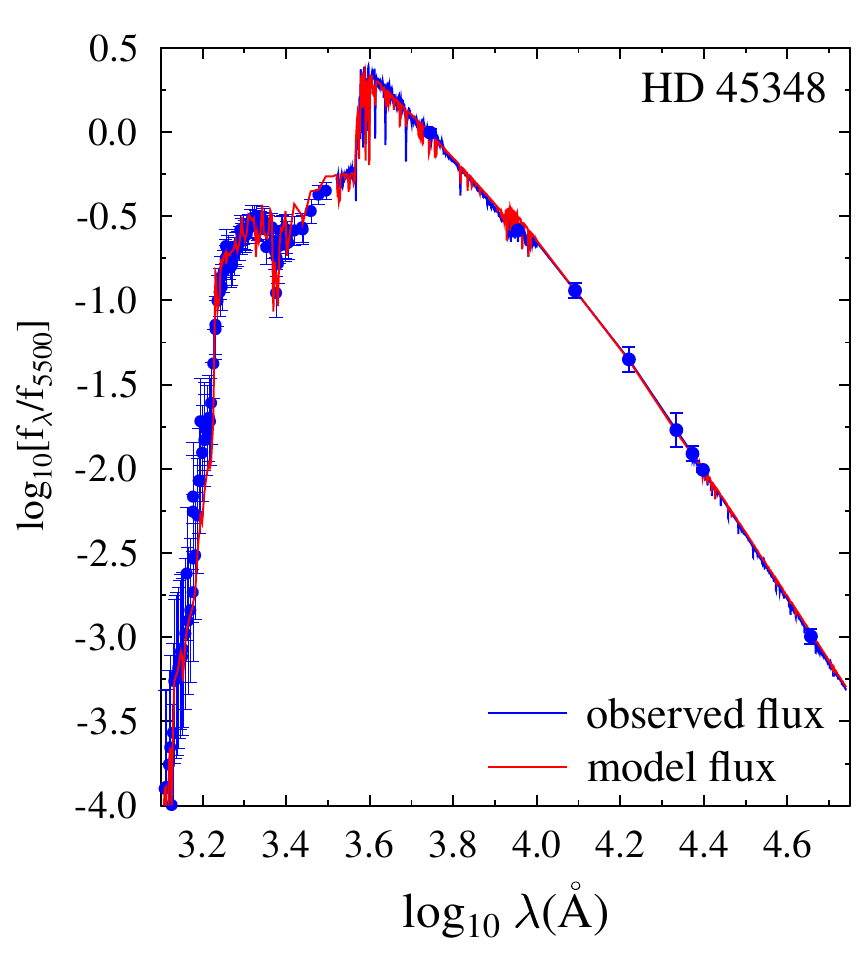}} 
\caption{\label{fig:SED} Best model SED obtained for Canopus (red), superimposed on the un-reddened observed SED (blue), with corresponding uncertainties. The observed and modeled fluxes are normalized to the respective fluxes in the visible at $\lambda=5500$~\AA\, ($\log 5500 = 3.74$). See details in Sects.~\ref{sed_obs} and \ref{zorec_params_from_sed}.}
\end{figure}

The missing spectral ranges, far-UV (below $1420$~\AA) and far-IR (beyond 160~$\mu$m), were completed with models adjusted at each iteration step of the effective temperature ($T_{\rm eff}$) and surface gravity ($\log g$) parameters (see Sect.~\ref{zorec_params_from_sed}). These domains contribute to the stellar bolometric flux with only $\sim1.6\times10^{-3}$~\% and $\sim2.8\times10^{-5}$~\%, respectively. The observed de-reddened SED of Canopus is given in Fig.~\ref{fig:SED}, together with our best-fit model (stellar and extinction parameters), which is described in the next section.

\subsection{Stellar parameters from the SED \label{zorec_params_from_sed}}

The fitting with model atmospheres of the Canopus SED is based on the simulated annealing iteration method \citep{Metropolis1953}. Using the interstellar medium (ISM) absorption law of \citet{Cardelli1989_v345p245}, the iterated primary parameters and their respective starting values are: $E(B-V)$ ($0.0$~mag), $R_{\rm V}$ ($3.1$), $T_{\rm eff}$ ($7314$~K), and $\log g$ ($1.82$~dex).

The starting values for $T_{\rm eff}$ and $\log g$ are averages of several independent determinations found in the literature \citep{przy1974,Kovtyukh2007_v378p617}, corresponding to a spectral MK type A9II-F0Iab \citep[e.g., SIMBAD-CDS,][]{Ayres2018_v854p95}, which place Canopus in the HR diagram midway between the blue and red supergiants. This led us to use the PHOENIX library of stellar local thermodynamic equilibrium (LTE) synthetic models for extended atmospheres with solar metallicity \citep{huss2013}. These models cover the wavelength range from $0.05$~$\mu$m to $5.0$~$\mu$m.

The iteration of fundamental parameters was done in two-step series. The first step was to determine the pair $(E(B-V),R_{\rm V})$ for a given pair $(\Teff, \log g)$ and the second to infer $(\Teff, \log g)$ using the previously obtained $(E(B-V),R_{\rm V})$. We went from one iteration series to the other as many times as required to minimize the chi-square computed over the entire SED as

\begin{equation}\label{eq:chi2_flux}
\chi^2_\mathrm{SED} = \sum_i\left(\frac{f^{\rm mod}_i-f^{\rm obs}_i}{\sigma_i}\right)^2 \,\,\, ,
\end{equation}

\noindent where $f^{\rm mod}_i=f^{\rm model}_{\lambda_i}(T_{\rm eff},\log g)$ and $f^{\rm obs}_i=f^{\rm observed}_{\lambda_i}(E(B-V),R_{\rm V})$ are, respectively, the modeled and the observed ISM extinction-corrected fluxes, with uncertainties $\sigma_i$ (see Sect.~\ref{sed_obs}).



\subsubsection{Fitting with a single $\Teff$ model}

Adopting the standard approach, we initially sought to fit the entire SED with a single effective temperature model. This first attempt, however, revealed that models do not reproduce all spectral domains well simultaneously. We found that while a model with a given $\Teff$ could produce a satisfying representation of the visible region, it did not simultaneously match the UV and IR. 

More specifically, the UV region can be modeled with $\Teff\simeq 7750$~K and $\log g\simeq1.55$~dex (with $E(B-V)\simeq 0.092$~mag), but this leads to a poor match of the IR flux. On the other hand, a good representation of the IR spectral range is attained with $\Teff\simeq 7290$~K and $\log g\simeq1.78$~dex (with $E(B-V)\simeq 0.004$~mag), but this produces an under-evaluated UV flux. A more complex model thus seems necessary for reproducing the entire SED of Canopus.

\subsubsection{Fitting with a two-$\Teff$ model \label{two_teff_model}}

A better fit of the entire SED is attained with two $\Teff$ and a single $\log g$. The model fluxes at each wavelength $\lambda$ are composed as 
\begin{equation}\label{eq:fllux_two_teff}
f^{\rm mod}_{\lambda}=(1-a)\times f^{\rm mod}_{\lambda}(T_{\rm eff_1},\log g)+
a\times f^{\rm mod}_{\lambda}(T_{\rm eff_2},\log g) \,\,\, ,
\end{equation}
\noindent where $a$ is a new iterated parameter that represents the fraction of the stellar surface radiating according to $T_{\rm eff_2}$. The reference effective temperature of the star for this two-$\Teff$ model is then
\begin{equation}\label{eq:global_teff}
\Teff=\left[(1-a)\times T^4_{\rm eff_1}+a\times T^4_{\rm eff_2}\right]^{1/4} \,\,\, .
\end{equation}

The best fit of this two-$\Teff$ model can reproduce the entire un-reddened observed SED, as shown in Fig.~\ref{fig:SED}. The corresponding model parameters are given in Table~\ref{tab:best_params_sed}, where other parameters for Canopus (mass $M$, angular diameter $\diameter$, age) are also given, the determination of which is described in the next section. The uncertainties affecting the fitted parameters listed in Table~\ref{tab:best_params_sed} were calculated using a Monte Carlo simulation. For the observed fluxes we adopted the uncertainties of the SED setting. The tested specific extinctions are for $2.8\lesssim R_{\rm V}\lesssim 3.3$, and the color excess $E(B-V)$ is given by a Gaussian distribution with dispersion $\epsilon_{\rm E(B-V)}=\pm0.06$ mag. 


%
\begin{table*}[ht!]
\centering
\caption{\label{tab:best_params_sed} Fundamental parameters of Canopus obtained from the best fit to the un-reddened SED.}
%
\begin{tabular}{*{2}{l}}
\toprule
\multicolumn{2}{c}{\textbf{Parameters derived from the SED model-fit}} \\
\toprule
Best-fit parameters from the two-$\Teff$ model & Related parameters \\
\midrule
$E(B-V) = 0.084$ mag, $R_{\rm V} = 3.1$ & $A_{\rm V}=0.26$~mag\\
$\log L/\Lsun = 4.221\pm0.018$ dex & $f_{\rm bol} = (59.22\pm2.45)\times10^{-6}$ erg\,cm$^{-2}$\,s$^{-1}$ \\
$T^{\rm mod}_{\rm eff_1} = 7800\pm164$ K,  $T^{\rm mod}_{\rm eff_2} = 7500\pm158$ K , $a = 0.49$ & $T^{\rm mod}_{\rm eff} = 7657\pm161$ K \\
$\diameter_{\rm Ross} = 7.19\pm0.34$ mas & $(R/\Rsun)_{\rm Ross} = 73.3\pm5.2$ \\
$\log g^{\rm mod} = 1.70\pm0.05$ dex & $(M/\Msun)_{\rm Ross} = 9.81\pm1.83$  \\
\toprule
\multicolumn{2}{c}{\textbf{Parameters derived using stellar evolution models}} \\
\multicolumn{2}{c}{(Values supposing Canopus on the blue loop phase, toward the blue or the red supergiants)} \\
\toprule
$\mathbf{\Omega/\Omega_{\rm c}=0.0}$ $\mathbf{(V/V_{\rm c}=0.0)}$ & $\mathbf{\Omega/\Omega_{\rm c}=0.95}$ $\mathbf{(V/V_{\rm c}\simeq0.81)}$  \\
\midrule
$(M/\Msun)_{\rm evol} =$ (9.64 or 9.63) $\pm 1.42$ & $(M/\Msun)_{\rm evol} =$ (9.26 or 9.25) $\pm 1.40$ \\
age$_{\rm evol}$ = (24 or 25) $\pm 3$ ~Ma & age$_{\rm evol}$ = (33 or 34) $\pm 4$ ~Ma \\

%
%
\bottomrule
\end{tabular}
\tablefoot{
Fundamental parameters of Canopus obtained from the best-fit to the un-reddened SED using the two-$\Teff$ model (upper part of the table) and from stellar evolution (Geneva) models (lower part). The parameters measured directly from $\log L/\Lsun$, global $T^{\rm mod}_{\rm eff}$ (two-$\Teff$ model; Eq.~\ref{eq:global_teff}), and $\log g$, called "Rosseland" parameters, are noted with the subindex "Ross." The parameters in the right column are related mainly (but not exclusively) to those in the left column on the same line by the adopted distance or the equations defined in Sects.~\ref{zorec_params_from_sed} and \ref{mass}. The parameters estimated with evolutionary models, noted with the subindex "evol," correspond to two ZAMS velocity ratios.  The Hipparcos parallax \citep[$\pi=10.55\pm0.56$~mas;][]{vanLeeuwen2007_v474p653} was used to relate the bolometric luminosity, $L$,  the bolometric flux, $f_{\rm bol}$, and the angular and linear sizes. See the discussion in Sect.~\ref{concordance_interf_sed}.
}
\end{table*}

\subsection{Mass, evolutionary status, and age estimates \label{mass}} 

The estimations of the Canopus mass $M$ given in Table~\ref{tab:best_params_sed} were obtained in two ways. The first was from the best-fit two-$\Teff$ model, through the iterated $\log g$ parameter and the radius, $R/\Rsun=(L/\Lsun)/(\Teff/T_{\rm eff_{\sun}})^4$, so $M/\Msun=(g/g_{\sun})(R/\Rsun)^2$. 

The second way was obtaining the mass from the Geneva evolutionary tracks for solar metallicity $Z=0.014$ \citep{geor2013} by adopting as entry parameters the measured $\log L/\Lsun$ and global $\Teff$ (Eq.~\ref{eq:global_teff}). As the internal rotational history of Canopus is unknown, we selected evolutionary tracks for angular velocity ratios at the zero-age main sequence (ZAMS) ($\Omega/\Omega_{\rm c}=0$ and $0.95$) to get a rough estimate of possible effects induced by rotation on the inferred stellar parameters. The $\Omega/\Omega_{\rm c}=0.95$ corresponds to a linear rotation velocity to critical velocity ratio of $V/V_{\rm c}\simeq0.8$ and an equatorial centrifugal to gravity acceleration ratio of $\eta\simeq 0.6$, which represent an object that is rapidly rotating in the ZAMS but still far from critical rotation. The standard deviations of masses and ages interpolated in the evolutionary tracks were obtained according to a Monte Carlo simulation of the uncertainties that affect the entry parameters.

\begin{figure}
\centerline{\includegraphics[width=\hsize, scale=1.0]{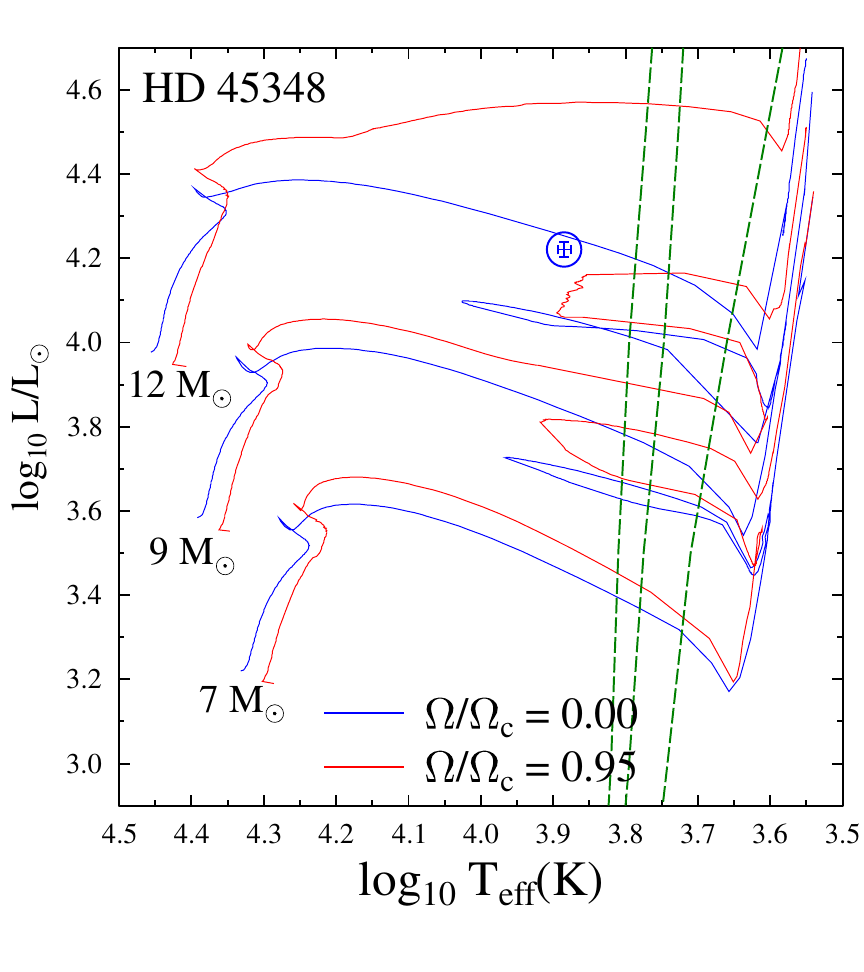}} 
\caption{\label{fig:Canopus_position_HR_diag} Position of Canopus in the HR diagram, indicated by the circle with the corresponding error bars within. Evolutionary tracks are from the Geneva models for solar metallicity ($Z=0.014$) and for masses similar to that of Canopus \citep{geor2013}. The colors of the tracks correspond to models without ($\Omega/\Omega_{\rm c}=0$; blue lines) and with rotation ($\Omega/\Omega_{\rm c}=0.95$; red lines) in the ZAMS. The dashed green lines represent the average crossing limits in the pulsation instability region of Cepheids \citep{and2016}. Our results indicate that Canopus is most probably located close to the hotter limit of the blue loop phase.}
\end{figure}

The ($\log L/\Lsun,T_{\rm eff}$) and radius obtained in our joint analysis of interferometry and SED, together with evolutionary models, allow the evolutionary status and age of Canopus to be estimated. These results exclude the possibility of Canopus being in the H shell burning stage (Hertzsprung gap) because the inferred mass would be larger by $\simeq2\Msun$ and $\log g=1.80$~dex, which conflicts with the "Rosseland" parameters in Table~\ref{tab:best_params_sed} obtained from the fit of the SED. Moreover, the apparent diameter would be a factor of $\gtrsim10$ too small ($\diameter\lesssim 0.7$ mas). The remaining possibility is that Canopus is presently in the blue loop region of the HR diagram, as illustrated in Fig.~\ref{fig:Canopus_position_HR_diag}. The associated age estimation is given in Table~\ref{tab:best_params_sed} for the two selected angular velocity ratios at the ZAMS.

Due to observational uncertainties, we can still ask whether the star is in the leftward or rightward lane of the blue loop. The masses and ages estimated from evolutionary models given in Table~\ref{tab:best_params_sed} correspond to the leftward lane. However, differences between rightward and leftward estimates are only $\sim-0.1\Msun$ and $\sim+1$~Ma 
for masses and ages, respectively.

\section{Discussion \label{discussion}}

\subsection{Comments on the two effective temperatures \label{two_teff_discuss}}

Two possible explanations support the need to use two temperatures to fit the whole SED. First, being in the blue loop region of the HR diagram, Canopus may have undergone evolutionary stages with great upheavals in its envelope that were produced by extended external convection movements and may have experienced phases of periodic changes of physical conditions in the external layers when crossing the pulsation instability strips. If rather large-scale movements still survive \citep{Strassmeier1998_v154p257,Gray1989_v341p421}, the temperatures in the granules and in the inter-granules will be different, such that the layers of constant Rosseland optical depth $\tau_{\rm Ross}=2/3$ are corrugated \citep{magic2013} and the emergent radiation field cannot reflect a single effective temperature.

The second is the failure of strict LTE models of extended atmospheres to represent the observed SED. Moreover, non-LTE models not only produce enhanced fluxes in the far-UV as compared to those predicted in LTE models \citep{mih78,auf99,haus99}, they are also strongly sensitive to the specific nature of the physical-structure-perturbed stellar atmospheres, as may be the case of Canopus. If so, this would possibly require 3D radiation transfer models that are not at our disposal.

\subsection{\label{compare_Neilson_SAtlas} Measured LD profiles compared to predictions from the SATLAS stellar atmosphere model}

We now compare the LD intensity profiles obtained from our analysis of PIONIER data to theoretical profiles computed with the SATLAS code \citep{Lester2008_v491p633}, a spherical version of the open-source model atmosphere program ATLAS from R.~Kurucz. SATLAS provides intensity profiles in spherically symmetric geometry, which is more realistic than plane-parallel models, particularly considering the low surface gravity of Canopus.

\begin{figure*}[ht!]
\centerline{\includegraphics[trim=3.8cm 1.2cm 4.6cm 2.0cm, clip,  width=0.85\hsize]{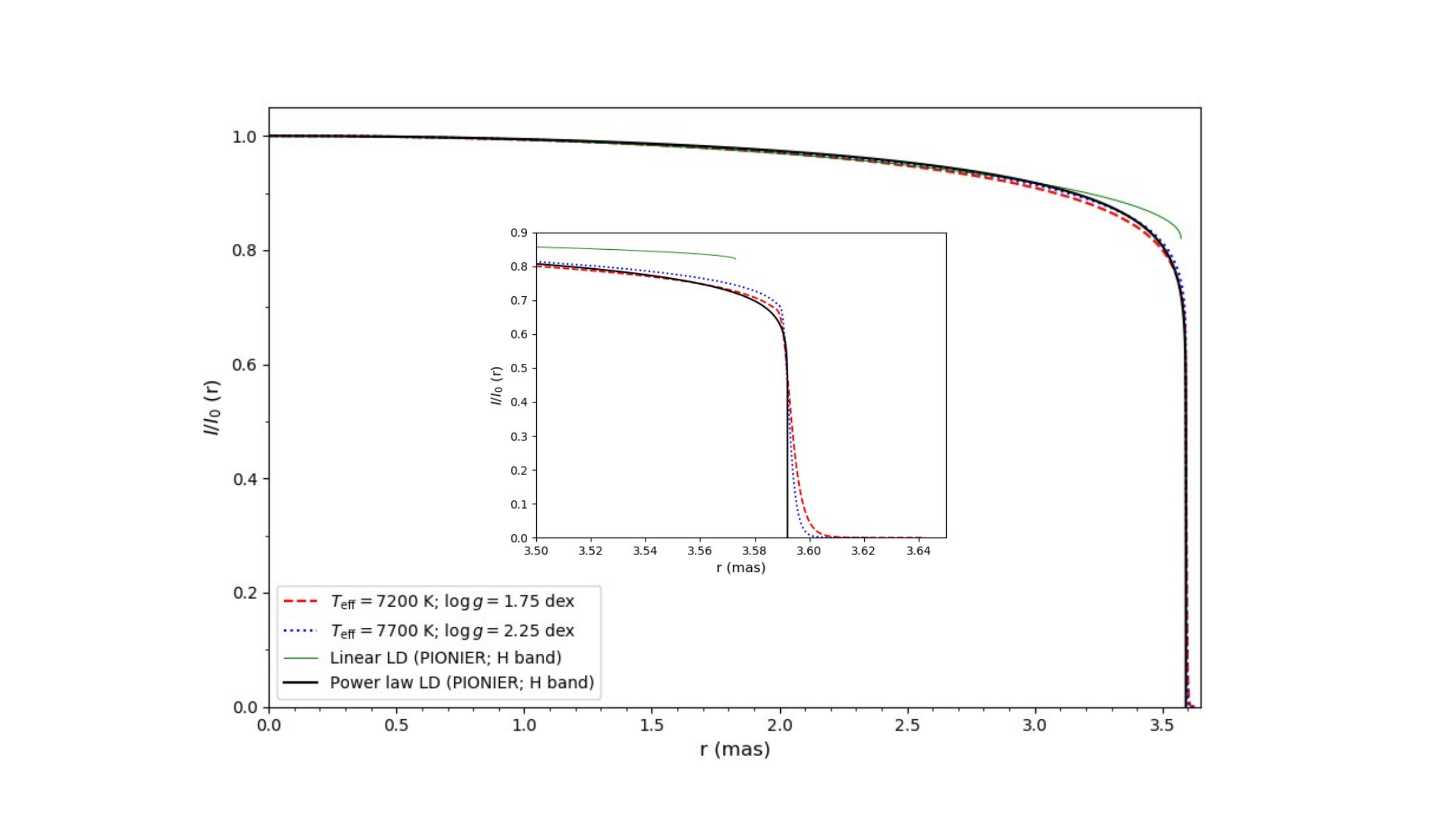}}
\caption{\label{fig:PIONIER_vs_SATLAS} Comparison between measured and modeled normalized intensity profiles, $I(r)/I(r=0)$, as a function of the angular radius $r$ (from 0 to $0.5$ of the angular diameter H). The interferometrically measured profiles are shown in plain lines for the linear (green) and power-law (black) LD models, fitted to the PIONIER data (see Tables~\ref{tab:ld_models} and \ref{tab:emcee_results_pionier}). The dashed red and dotted blue profiles correspond to selected SATLAS stellar atmosphere models computed in spherically symmetric geometry in the H band \citep{Lester2008_v491p633, Neilson2013_v554pA98}. The selected models correspond to a $10\Msun$ model, with $\Teff$ (blue) or $\log g$ (red) being closest to the measured values. The measured power-law LD and the SATLAS profiles show a general good agreement, even though they do not exactly match the measured Canopus parameters. A small discrepancy is mostly seen only very close (above 97\%) to the stellar limb (see the zoomed-in plot in the inset). The precision of present interferometric data does not allow  such tiny differences to be distinguished among these intensity profiles.}
\end{figure*}

Figure~\ref{fig:PIONIER_vs_SATLAS} shows the normalized intensity profiles for the measured linear and power-law LD parameters (Table~\ref{tab:emcee_results_pionier}) together with two SATLAS theoretical H-band profiles, selected from the grid provided by \citet{Neilson2013_v554pA98}. The two selected SATLAS profiles correspond to $10\Msun$ models, with $\Teff$ or $\log g$ close to those of Canopus. There are no models with both $\Teff$ and $\log g$ close to the measured values. For the comparisons we made the intensity profile dropouts seen in the SATLAS models coincide with the stellar limb defined by our best-fit power-law LD model.

The theoretical SATLAS profiles are in good agreement with the power-law model computed with the measured LD coefficient ($\alpha_\mathrm{H}=0.1438$). The measured linear LD model presents a poorer agreement with the SATLAS theoretical computations, especially close to the limb ($r$ above $\sim90\%$ of the stellar radius).


\subsection{Concordance between interferometry and SED analysis \label{concordance_interf_sed}}

In our analysis of the interferometric and SED data of Canopus, several distinct fundamental parameters were measured, as well as one identical parameter, the angular diameter, where we found that $\diamH$ (power-law LD from PIONIER) and $\diameter_{\rm Ross}$ (SED) agree well within their uncertainties. Thanks to this agreement, we could then use the $\simeq10$ times more precise $\diamH$ to further improve some parameter values and errors. 

From our best-fit power-law LD $\diamH$ given in Table~\ref{tab:emcee_results_pionier} and from the Hipparcos parallax, we straightforwardly computed the Canopus radius: $R=0.5 \diamH d=73.2 \pm 3.9 \Rsun$. The three uncertainties (statistical and systematic from $\diamH$ and the error in the distance) were quadratically added to obtain the radius error. The interferometric radius is thus in very good agreement with the value from the SED analysis (see Table~\ref{tab:best_params_sed}), and it is also more precise, as expected. We note that the error in the interferometric radius is dominated by the uncertainty of the Canopus distance, so future, more precise distance measurements (for instance from Gaia) would directly lead to an even lower radius error using the interferometric $\diamH$ measured in this work. 

We can also somewhat improve our estimation of the Canopus global $\Teff$ through the relation \citep[e.g.,][]{Jerzykiewicz2000_v50p369}: 
\begin{equation} \label{eq:teff_fbol_theta}
\Teff = 7401.6 f_{\rm bol}^{\frac{1}{4}} \diameter^{\frac{1}{2}} \,\,\, ,
\end{equation}
with $\diameter$, $\Teff$, and $f_{\rm bol}$ given in mas, K, and $10^{-6}$ erg\,cm$^{-2}$\,s$^{-1}$, respectively. From our bolometric flux, $f_{\rm bol}$, and $\diamH$ we obtain $\Teff=7661\pm81$~K. This estimate agrees nicely with the SED measurement (Table~\ref{tab:best_params_sed}), as expected, and has a lower uncertainty (by a factor of $\sim2$) thanks to the use of the more precise $\diamH$.


\subsection{Comparison of our results with some previous works}

We compare here some physical parameters of  Canopus that we measured with previously reported values from selected works. These comparisons are not intended to be exhaustive but are presented to point out some important agreements and discrepancies, as well as to highlight the improvements provided by the present work.

\subsubsection{Angular diameter and limb darkening}

Since Canopus is a very bright star, its angular diameter, $\diameter$, has been estimated by many researchers using different techniques, both indirectly (e.g., from spectral fluxes or lines) or directly from OLBI (amplitude or intensity interferometry). A compilation of many measured $\diameter$ is given by \citet[][Table 5]{Cruzalebes2013_v434p437}, with values ranging from 5.9 to 7.2~mas and with relative uncertainties of $\sim2-12\%$. The best-fit interferometric $\diameter$ measured in this work (Table~\ref{tab:emcee_results_pionier}) presents a relative uncertainty of $\approx0.4\%$, which is significantly more precise than all these previous results. In particular, in the present work the relative uncertainty on $\diameter$ and on the LD coefficient are improved, respectively, by a factor of $\gtrsim 5$ and $\sim15-25$ with respect to our previous work on Canopus \citep{DomicianodeSouza2008_v489pL5}, where we analyzed a different and lower S/N AMBER data set, as previously mentioned.

Regarding values of $\diameter$ derived using indirect methods, we would like to note two works that provide estimations in close agreement with our result, though, as expected, with much higher uncertainties: \citet{Blackwell1977_v180p177} obtained $7.08\pm0.19$~mas from several IR measurements, and \citet{Decin2003_v400p695} obtained $7.22\pm0.42$~mas from absolutely calibrated ISO spectra from the Short Wavelength Spectrometer (SWS). These agreements become even more noticeable if one considers that these authors estimated $\diameter$ for the star $\alpha$~Cen~A as well, obtaining values that are also in good agreement with the direct and more precise measure from \citet{Kervella2017_v597pA137}, who used PIONIER data of quality similar to those used in the present work.

\subsubsection{$\Teff$ and bolometric flux}

As mentioned in Sect.~\ref{zorec_params_from_sed}, several values of $\Teff$, $\log g$, and other parameters have been reported in the literature. In addition to the aforementioned works we can also cite, for example: \citet{Soubiran2016_v591pA118}, \citet{Cruzalebes2013_v434p437}, \citet{DomicianodeSouza2008_v489pL5}, \citet{Smiljanic2006_v449p655}, \citet{Decin2003_v400p695}, \citet{Jerzykiewicz2000_v50p369}, \citet{Achmad1991_v249p192}, \citet{Desikachary1982_v201p707}, and the references therein, which span many decades of publications. 

In these works the bulk of the reported $\Teff$ lies roughly in the range $7200 \lesssim \Teff \, \mathrm{(K)} \lesssim 7575$. From this $\Teff$ range and considering our interferometric angular diameter, $\diameter$ (Table~\ref{tab:emcee_results_pionier}), one gets from Eq.~\ref{eq:teff_fbol_theta} that $46.2 \lesssim f_{\rm bol} \lesssim 56.6$, in units of $10^{-6}$ erg\,cm$^{-2}$\,s$^{-1}$. It is interesting to note that even the lowest $f_{\rm bol}$ value in this range is already close to most reported values \citep[e.g.,][in addition to the already cited references]{Code1976_v203p417, Blackwell1980_v82p249, McWilliam1991_v101p1065, Smalley1995_v293p446}. 

There is thus an inconsistency between these reported measurements of $\Teff$ and $f_{\rm bol}$ if we take our interferometric $\diameter$ as a realistic face value for Canopus. This inconsistency does not exist in the present work since there is a concordance (see Sect.~\ref{concordance_interf_sed}) between the interferometrically obtained $\diameter$ and the parameters from the SED analysis ($\Teff$ and $f_{\rm bol}$ in particular).

\subsubsection{Mass and $\log g$}

The references mentioned in the last subsection also report values of surface gravity and mass, respectively, in the ranges $1.2 \lesssim \log g \, \mathrm{(dex)}\lesssim 2.4$ and $7.0 \lesssim M/\Msun \lesssim 12.8$. These wide ranges illustrate the difficulty inherent to the estimation of these parameters in single stars. Our estimates lay well within these ranges, in agreement with several works, in particular with the evolutionary-based values from \citet{Smiljanic2006_v449p655} and \citet{Jerzykiewicz2000_v50p369}.



\subsection{\label{activity_spots} Activity, temporal variability, and spots}



Canopus presents high energy emission in the UV and X-ray, the physical origin of which is not completely understood:
(i) a magnetic field of several hundred gauss, measured from Zeeman shifts on UV spectral lines, associated with periodic variations of a few days to a few weeks \citep[e.g.,][]{Weiss1986_v160p243, Rakos1977_v56p453, Bychkov2005_v430p1143, Bychkov2009_v394p1338}; (ii) far-UV emission lines showing non-symmetrical bisector curves with reverse C-shapes, revealing the present of opacity effects and/or velocity fields \citet{Dupree2005_v622p629}; and (iii) a high X-ray luminosity $L_\mathrm{X}$ of a few $10^{30}$~erg\,s$^{-1}$ \citep[e.g.,][]{Vaiana1981_v245p163, Strassmeier1998_v154p257, Hunsch1998_v127p251, Testa2004_v617p508, Ayres2011_v738p120, Ayres2017_v837p14, Ayres2018_v854p95}.

All these compelling results show that Canopus has some atmospheric activity, including related surface inhomogeneities. On the one hand, this provides further justification for the use of our two-$\Teff$ model (Sect.~\ref{two_teff_model}) and, on the other hand, invites us to search for these inhomogeneities in our high angular resolution data. However, their presence is expected to be much subtler and hard to detect in the visible and IR because Canopus is not known to show strong surface inhomogeneities and/or time variations in these spectral domains.

\subsubsection{\label{intef_signature_spots} Interferometric signatures of photospheric inhomogeneities from a simple model fitting}

To investigate the presence of surface inhomogeneities on Canopus, we performed another \textit{emcee} model fitting on the VLTI data, considering a limb-darkened star with one UD photospheric spot as a first order representation of a global photospheric inhomogeneity. 

The adopted LD model is the power-law model presented in Table~\ref{tab:ld_models}. A UD photospheric spot is added to this LD model and is allowed to be located on the visible stellar disk. The following additional parameters are required to model this UD spot: the spot size, $S_\mathrm{s}$, relative to the stellar angular diameter, $\diameter$; the 2D position coordinates $x_\mathrm{s}$ and $y_\mathrm{s}$ relative to the center of the stellar disk; and the ratio $f_\mathrm{s}$ between the fluxes of the spot ($F_\mathrm{s}$) and the limb-darkened star ($F_\mathrm{LD}$). We consider $f_\mathrm{s}>0$, but it is expected to be very low, as discussed above. The fit was performed including a prior condition on $S_\mathrm{s}$, $x_\mathrm{s}$, and $y_\mathrm{s}$ to ensure that the UD spot is entirely contained inside the region delimited by the visible stellar disk, with $S_\mathrm{s}$ ranging from zero to $0.9\diameter$ (almost the whole star). Bandwidth smearing was not included in order to maximize the signal of the putative spot.

We fitted this combined model to the same PIONIER and AMBER (H and K) data used in Sect.~\ref{interf_emcee_results}. The \textit{emcee} model fitting could not converge to well-defined values for all parameters (multimodal distributions), especially for the AMBER data, which are noisier than the PIONIER ones. The best-fit angular diameter, $\diameter$, and power-law LD coefficient, $\alpha$, agree with the values in Table~\ref{tab:emcee_results_pionier}. For the AMBER data these two parameters are compatible with the results in Table~\ref{tab:emcee_results_amber} if we consider values associated with the minimum $\chi^2$ (not the median). For the corresponding UD spot parameters, our fitting procedure could only poorly determine the spot size and position. However, the fit shows that the PIONIER and AMBER data constrain the flux ratio (spot to LD star) to a low value, as expected: $f_\mathrm{s}= F_\mathrm{s}/F_\mathrm{LD} \sim 0.001-0.002$ (PIONIER) and $\sim0.003-0.005$ (AMBER). Although the contribution of the spot to the total flux is small, including it in the model leads to a slightly better fit quality, with a total $\chir$ that is about $5-15\%$ lower than those for the power-law LD model alone. The presence of low contrast inhomogeneities on Canopus seems therefore compatible with the interferometric data investigated in this work. 


Moreover, our results are in line with \citet{Cruzalebes2015_v446p3277}, who analyzed an independent set of interferometric Canopus data (AMBER in the K band at medium resolution) to search for signatures of departures from centrosymmetry. Using an approach distinct from ours, these authors also found that Canopus presents a real but marginal interferometric signature of inhomogeneity. The results of our present work, obtained from quality VLTI observations, also confirm our previous results regarding the presence of surface structures on Canopus \citep{DomicianodeSouza2008_v489pL5}, even though differences in the determined parameters exist (e.g., flux ratio), which is not surprising since the previously analyzed data are less numerous and present a lower S/N, as already mentioned.

Finally, we note as well that our flux ratio, $f_\mathrm{s}$, estimation for PIONIER is on the same order as the best estimation found by \citet{Neilson2014_v563pL4}, after converting their values of spot opening angle and temperature to the equivalent $f_\mathrm{s}$. They found that this level of spot flux fraction can explain the small period jitter observed in some Cepheids, which are not far from Canopus in the HR diagram.

Thus, the signature of surface inhomogeneities, suggested by our analysis, provides additional and independent observational evidence for the presence of some degree of activity and/or asymmetries on Canopus, also supporting the results from all the aforementioned studies.

\subsubsection{Image reconstruction \label{imaging}}

Our results from Sect.~\ref{intef_signature_spots}, suggesting the presence of weak surface inhomogeneities on Canopus, call for an attempt to reconstruct  interferometric images, especially because of the good $uv$-plane coverage of the VLTI data.

We performed a model-independent image reconstruction on the PIONIER data with MIRA\footnote{\href{https://cral-perso.univ-lyon1.fr/labo/perso/eric.thiebaut/?Software/MiRA}{Multi-aperture Image Reconstruction Algorithm}} software \citep{Thiebaut2008_v7013p70131I}. We considered only the 2014 observations (which are contemporaneous, i.e., obtained within a one-month time span) that essentially correspond to the same "view" of the surface of Canopus. Unfortunately, even with this precaution, we found that the reconstruction is ambiguous. The final solution depends on the choice of the regularization, on the regulation weight, and slightly on the starting point. Nevertheless, whatever the reconstruction made, the difference between the reconstructed images and the closest featureless model is $\lesssim1$\%. Thus, these faint features are real, although their exact geometry is unconstrained. 

We also attempted to perform image reconstruction with the AMBER data. The results are even less conclusive than for the PIONIER data, likely because of their lower quality, as discussed earlier in the text. These difficulties found in obtaining conclusive, \textit{bona fide} reconstructed images of Canopus are compatible with, and at the same time independently confirm, the very low image contrast associated with the weak spot flux found in our model fitting procedure.


\section{Conclusions \label{conclusions}}

In this work we have determined a new set of fundamental parameters of Canopus, presenting a mutual concordance between interferometric data from the VLTI (PIONIER and AMBER) and the SED (from the UV to the IR), built from several published observations, in particular recent good quality visible data.

Because of their good precision and accuracy, the PIONIER data provide the central reference of our interferometric analysis of Canopus. The PIONIER data, reaching the third visibility lobe of Canopus, provided a good constraint on LD laws with one coefficient. We have thus estimated the angular diameter and LD coefficient of Canopus for the first time with a relative precision of $\simeq0.4\%$ and $\simeq1\%$, respectively. These are, to our knowledge, the most precise measurements of these two parameters for Canopus, in particular constituting an improvement by a factor of $\gtrsim 5$ and $\sim15-25$, respectively, with respect to the precision achieved in our previous work \citep{DomicianodeSouza2008_v489pL5}.

We show that the power-law LD model is in good agreement with previous theoretical predictions from the SATLAS stellar atmosphere model (spherical symmetry). These results thus provide an invaluable observational validation for theoretical models in this region of the HR diagram, which corresponds to the yellow supergiants. Indeed, the precise diameter and LD measured in this work are invaluable observational constraints for realistic (3D) physical models, which still have difficulties in simulating the structure and evolution of such evolved, hot and luminous stars with relatively high $\Teff$ and low $\log g$. 

Moreover, since the precision of the measured angular diameter of Canopus is limited mainly by systematical uncertainties associated with instrumental wavelength calibrations, improvements of these calibrations may allow the precision of future interferometric measurements to be increased even further. Angular diameters with precision similar to that of this work ($\lesssim1\%$) are expected to be obtained for several hundred stars with the near-future Stellar Parameters and Images with a Cophased Array \citep[SPICA\footnote{\url{https://lagrange.oca.eu/fr/spica-project-overview}};][]{Mourard2018_v0701p1070120}, the new visible beam combiner for the Center for High Angular Resolution Astronomy (CHARA) interferometer. 


Based on previous works that report the presence of stellar activity on Canopus, we investigated this point in detail using the interferometric data, both with model fitting and imaging techniques, and find indications that some surface inhomogeneities can be present but at a very low contrast, preventing their precise characterization. Our analysis indicates a flux ratio (spot to LD star) of at most 0.005, and 0.002 if we consider only the most precise PIONIER data. These results also agree with other previous works. More precise interferometric observations would be necessary to provide additional constraints on these surface structures.

The SED analysis combined with predictions of stellar evolution codes allowed us to measure several fundamental parameters of Canopus: reddening, temperature, bolometric flux, luminosity, mass, gravity, radius, and age. This set of fundamental parameters is secured by the very good agreement in the angular diameter obtained from the SED and interferometric analysis.

Thus, the fundamental parameters of Canopus measured in this work constitute a careful balance of the different methodologies used, providing invaluable observationally based constraints to models of the stellar atmospheres and stellar evolution of evolved massive stars.

\begin{acknowledgements}
This research has made use of the Jean-Marie Mariotti Center (JMMC) services \texttt{OiDB}\footnote{\url{http://oidb.jmmc.fr}}, \texttt{OIFits Explorer}\footnote{\url{http://www.jmmc.fr/oifitsexplorer}}, and \texttt{AMHRA}\footnote{\url{https://amhra.jmmc.fr/}}, co-developed by CRAL, IPAG, and Lagrange/OCA. ADS acknowledges A. Chiavassa and L. Bigot from OCA for enlightening discussions. We thank the CNRS (France) for the GTO time (VISA-CNRS AMBER) allocated to this project. This research made use of the SIMBAD and VIZIER databases (CDS-Strasbourg astronomical Data Center, Strasbourg, France) and NASA's Astrophysics Data System. This research has made use of the NASA/IPAC Infrared Science Archive (IRSA), which is funded by the National Aeronautics and Space Administration and operated by the California Institute of Technology.


\end{acknowledgements}

\bibliographystyle{aa} 
\bibliography{Canopus_2021} 

\newpage
\begin{appendix}

\section{Interferometric visibility for a general analytical limb-darkening law \label{app:analytical_vis}}

The analytical functions presented in Table~\ref{tab:ld_models} are special cases of a general form of a radial intensity profile (LD model) composed of powers of $\mu$:
\begin{equation}\label{eq:Imu}
\frac{I{_\lambda}(\mu)}{I{_\lambda}(1)}= \sum_{j=1}^{n} a_j \mu^{b_j}  \,\,\, ,
\end{equation}
where, as already mentioned, $\mu (= \cos(\theta))$ is the cosine of the angle $\theta$ between the direction perpendicular to the stellar surface and the observer's direction.

The corresponding complex visibility can be obtained from the Hankel transform of this intensity profile \citep[see, for example,][]{Hestroffer1997_v327p199, DomicianodeSouza2003_PhD}:
\begin{equation}\label{eq:V_hankel}
V = \frac{  \sum_{j=1}^{n} \frac{a_j}{2} \Gamma({\nu_j}) \displaystyle  2^{\nu_j} \frac{J_{\nu_j}(z)}{z^{\nu_j}}}{\sum_{j=1}^{n} \frac{a_j}{2{\nu_j}}} = \frac{  \sum_{j=1}^{n} \frac{a_j}{2{\nu_j}} \Gamma({\nu_j}+1) \displaystyle  2^{\nu_j} \frac{J_{\nu_j}(z)}{z^{\nu_j}}}{\sum_{j=1}^{n} \frac{a_j}{2{\nu_j}}}  \,\,\, ,
\end{equation}
where $a_j$, $b_j$, and ${\nu_j}(=b_j/2+1)$ are real numbers and functions of the wavelength $\lambda$ (not explicitly indicated here). Also, $\Gamma$ is the gamma function, and $J_\nu$ is the Bessel function of first kind and of order ${\nu_j}$. The dimensionless variable $z$ is given by 
\begin{equation}
z = \frac{\pi \diameter \Bp}{\lambda}  \,\,\, ,
\end{equation}
where $\Bp$ is the projected baseline, $\lambda$ is the effective wavelength, and $\diameter$ is the stellar angular diameter.

\section{Bandwidth smearing \label{band_smear}}


We included the bandwidth smearing effect in the modeling of the visibility amplitudes $\lvert V( \lambda ,u,v) \rvert$ by dividing each observational spectral bin into $N$ sub-bins, computing the individual $\lvert V( \lambda _{j} ,u_{j} ,v_{j}) \rvert$ in each sub-bin $j$, and finally calculating the average, weighted by the flux $F(\lambda _{j}) $, which is expected to be received by the sub-bin:
\begin{equation}
\lvert V( \lambda ,u,v) \rvert=\frac{\sum ^{N}_{j=1} F( \lambda _{j}) \lvert V( \lambda _{j} ,u_{j} ,v_{j}) \rvert}{\sum ^{N}_{j=1} F( \lambda _{j})}
.\end{equation}

The quantity $F|V|$ is the correlated flux. In the case of the present VLTI observations with a spectral resolution of a few tens, $F(\lambda _{j})$ can be considered constant within the bin, so the calculations including bandwidth smearing are simplified to:
\begin{equation}
\lvert V( \lambda ,u,v) \rvert=\frac{\sum ^{N}_{j=1} \lvert V( \lambda _{j} ,u_{j} ,v_{j}) \rvert}{N}
.\end{equation}

In this work we used $N\sim20-30$ sub-bins, determined such that the numerical spectral resolution for the model calculations $\lambda/\delta\lambda$ is 1000.


%
%

%
%

\section{Power-law LD model fitting to the AMBER H and K data \label{best_fit_power_law_AMBER}}

Based on the best-fit model obtained from the PIONIER data analysis, we performed an \textit{emcee} model fitting of the power-law LD model (with bandwidth smearing) to the AMBER data.

Before performing this model fitting we found it necessary to artificially increase the $V^2$ uncertainties for the few points close to the first and second $V^2$ minima in order to ensure a good convergence of \textit{emcee}. After some tests we decided to set a lower limit to the data points that present underestimated errors. The chosen limit value was $\sigma V^2_\mathrm{min} = 0.0001$, estimated from the typical $V^2$ errors found in the vicinity of the $V^2$ minima. This $\sigma V^2_\mathrm{min}$ was added quadratically to the original $V^2$ errors when the observed $V^2$ values corresponded to correlated magnitude differences, $\Delta \mathrm{mag}$ ($=-1.25 \log V^2$) $\geq4$ mag (i.e., $V^2 \leq 0.00063$). This roughly corresponds to the performances of AMBER in terms of measurements of magnitude differences.

\begin{figure*}[ht!]
\centering
\subfloat[AMBER H band]{\includegraphics[clip, trim=2cm 0.5cm 16cm 0.6cm, width=0.48\textwidth]{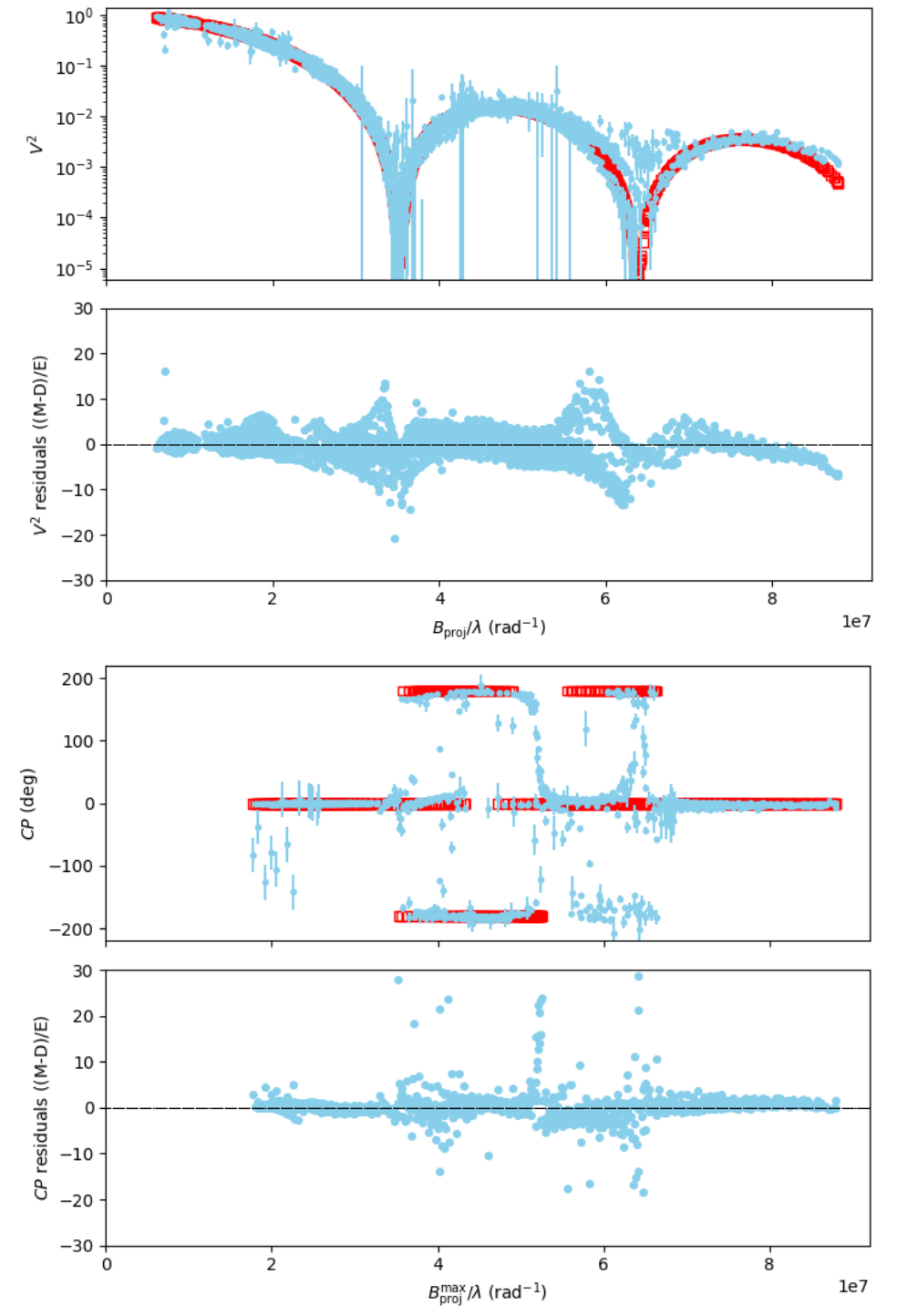}}%
\qquad
\subfloat[AMBER K band]{\includegraphics[clip, trim=2cm 0.5cm 16cm 0.6cm, width=0.48\textwidth]{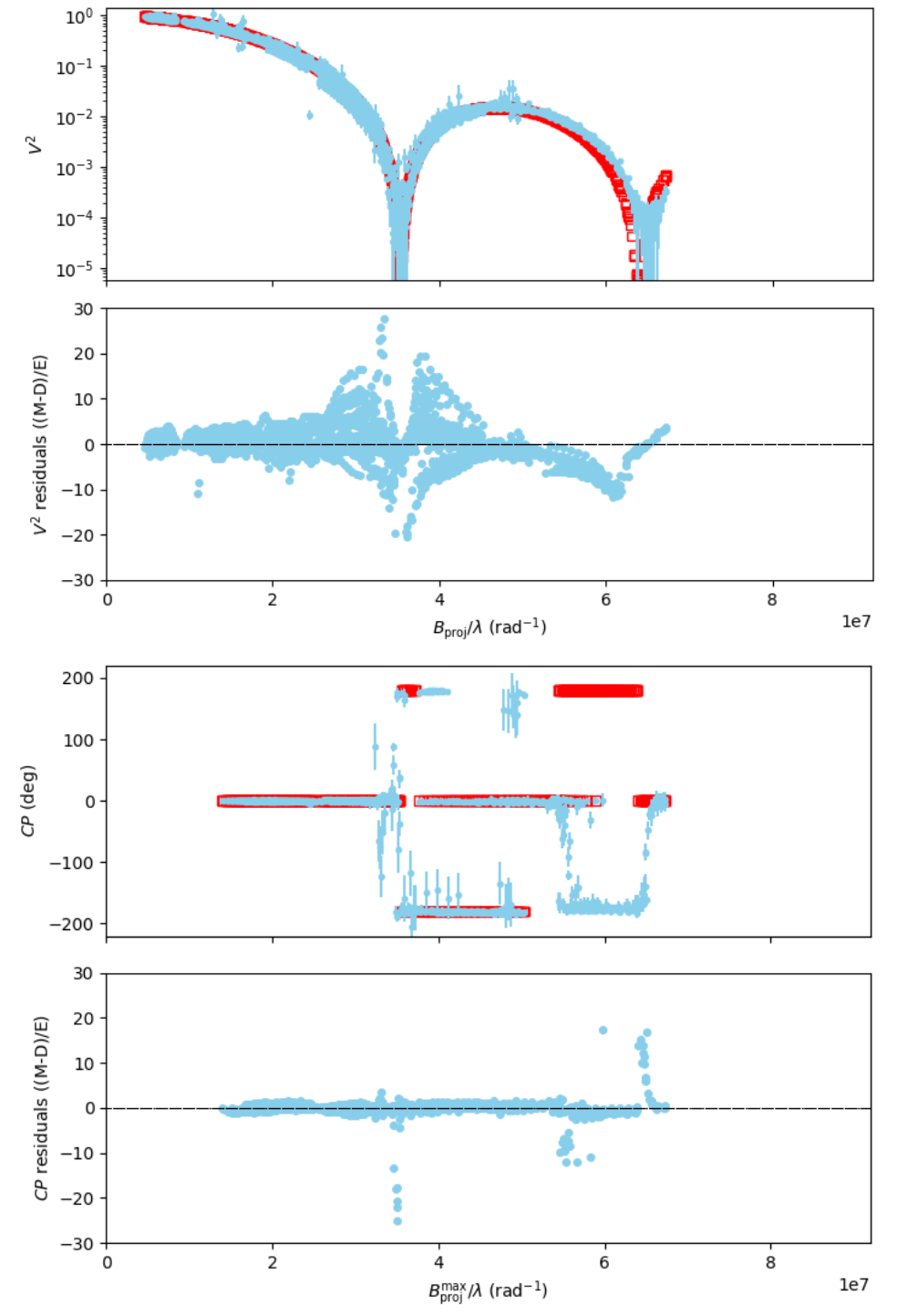}}
\caption{\label{fig:best_fit_power_law_AMBER} Canopus calibrated squared visibilities, $V^2$ ($\log$ scale), closure phases, $CP$, and errors as a function of the spatial frequency, observed with AMBER/VLTI (light blue) in the H (a) and K (b) bands. The red squares correspond to the best-fit-model observables obtained from an MCMC fit on the AMBER data using a power-law LD and including bandwidth smearing. The residuals of the fit are also shown.}
\end{figure*}

Table~\ref{tab:emcee_results_amber} presents the best-fit results (parameter values and uncertainties) derived from the fit of the power-law LD model to the AMBER H and K data. Figure~\ref{fig:best_fit_power_law_AMBER} shows the comparison of this best-fit model to the $V^2$ and $CP$ measured on Canopus. Although the model reproduces the interferometric observables fairly well, the AMBER data present a much higher dispersion when compared to the PIONIER results seen in Fig.~\ref{fig:best_fit_power_law_PIONIER}, which ultimately leads to lower-precision measurements.


%
\begin{table}[]
\centering
\caption{\label{tab:emcee_results_amber} Best-fit results for a power-law LD model, including bandwidth smearing, fitted to the Canopus AMBER low resolution (LR) data. }
\begin{tabular}{*{3}{c}}
\toprule
\multicolumn{3}{c}{\textbf{Power-law LD with bandwidth smearing}} \\
\toprule
\textbf{AMBER} & \textbf{Parameters} & $\chir$\\
\textbf{LR} &  & (total/$V^2$/$CP$) \\
\midrule
H band & $\diamH=7.31 \pm 0.05 \pm  0.15 $ mas & 12.3/7.7/4.6\\
       & $\alpha_\mathrm{H}=0.12 \pm 0.03 $ & \\
\midrule
K band & $\diamK=7.33 \pm 0.09 \pm 0.11 $ mas & 16.6/15.2/1.4 \\
       & $\alpha_\mathrm{K}=0.16 \pm 0.05 $ & \\
\bottomrule
%
%
%
\end{tabular}
\tablefoot{
Best-fit results for a power-law LD model (defined in Table~\ref{tab:ld_models}), including bandwidth smearing, fitted to the Canopus AMBER LR data: H band (3495 data points: 2619 $V^2$ and 876 $CP$) and K band (3911 data points: 2933 $V^2$ and 978 $CP$). The best-fit parameter values correspond to medians obtained from the histograms provided by \textit{emcee}. The second values correspond to statistical uncertainties. The third values given for the angular diameters $\diamH$ and $\diamK$ are the systematical errors associated with the instrumental wavelength calibration of AMBER (see Sect.~\ref{sfreq_precision}). We also provide the reduced chi-square, $\chir$, for the whole data set and for the $V^2$ and $CP$ data alone.
}
\end{table}
%

\end{appendix}

\end{document}